\documentclass[11pt,twoside]{article}
\usepackage{asp2010}

\resetcounters

\begin{document}

\title{Application of Binary Evolution in Astrophysical Studies}
\author{Zhongmu Li$^{1,2}$
\affil{
 $^{1}$National
Astronomical Observatories, the Chinese Academy of Sciences,
Beijing 100012, China\\
$^{2}$Institute for astronomy and history of science \& technology,
Dali University, Dali 671003, China\\
} }

\begin{abstract}
Binary stars evolve differently from single stars, thus binary
evolution is very useful for astrophysical studies. This paper
discusses the application of binary evolution in the studies of
stars, star clusters, galaxies, and cosmology. In particular, I
concentrate on the use of binary evolution in colour-magnitude
diagram and spectral energy distribution studies of star clusters
and galaxies.
\end{abstract}

\section{Binaries and Their Application in Astrophysical Studies}
Binary stars are very common in the universe. For example, the Milky
Way Galaxy includes more than fifty percent binary stars. A lot of
astrophysical processes and properties can be investigated via
binary stars. In fact, binary star is one of the most powerful tool
for astrophysical studies. First, binary star is a good tool for
studying many properties of stars. Usually, the luminosity,
acceleration of gravity, effective temperature, and colours of stars
in binaries are different from those of single stars. To investigate
the properties of stars, the effect of binary interaction and binary
evolution should be taken into account. Second, binary star is
useful for stellar evolution study. Because there are many
interactions between two components of a binary, the evolution of
binary stars are different from single stars. The evolution of
binary star has been the most important subject in studies about
star evolution. In Figure 1, a comparison of the evolution of binary
and single stars is shown. We see that in the binary and single
evolution modes, stars evolve differently. Note that the evolution
of stars are calculated via a rapid stellar evolution code (Hurley
et~al. 2002). Third, binary stellar evolution has an important role
in studies of star clusters and galaxies. The reason is that star
clusters and galaxies contain many binaries, and their integrated
properties are related to binary evolution. Especially, when
studying the stellar populations of star clusters and galaxies,
binary evolution can affect the results obviously. Binary and single
star evolution usually give different colour-magnitude diagrams
(CMDs) and integrated spectral energy distributions (SEDs) for star
clusters and galaxies. Fourth, binary evolution is helpful in
cosmological studies. One method to constrain the age of the
universe is to measure the correct ages of old globular star
clusters and then give a lower-limit for the universe's age. The
result is actually related to binary evolution, because binary
evolution lead to different integrated specialties of stellar
populations. In addition, binary evolution may also affect the
studies like dark matter and star formation.
\begin{figure} 
  \includegraphics[angle=-90,width=0.9\textwidth]{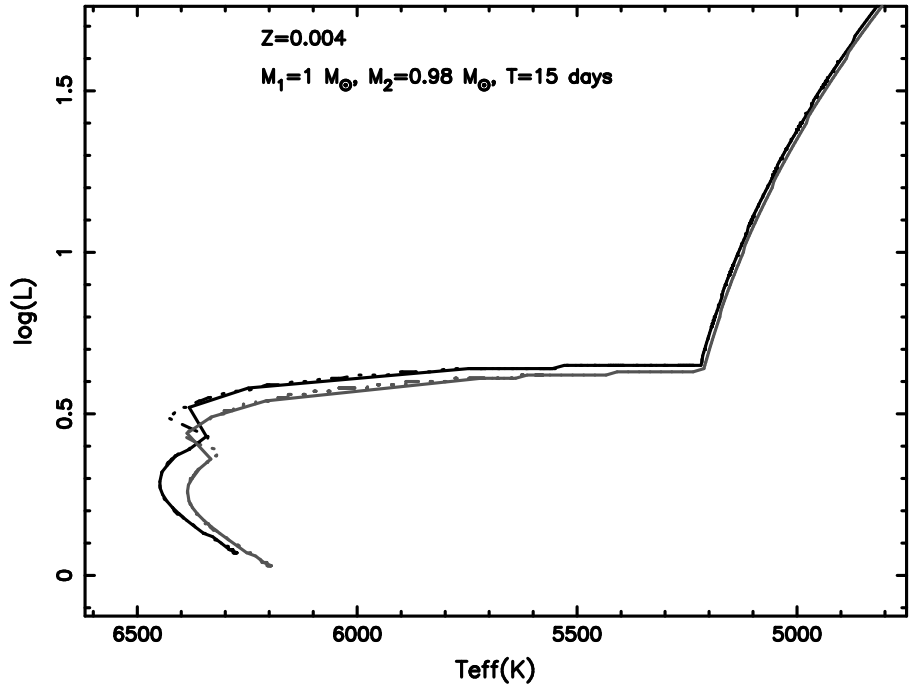}
  \caption{Comparison of single and binary stellar evolution of two stars with a metallicity of $Z$ = 0.004.
  Two stars with 1.0 (star 1) and 0.98 (star 2) solar mass are investigated.
  Eccentricity of the binary is randomly set to be 0.7, and an orbital period of 15 days is taken.
  Solid lines are for single evolution of two stars, while dash-dot-dot-dot lines for binary evolution.
  Stars 1 and 2 are shown in black and gray, respectively.}
\end{figure}

As an important part, this paper will show some new results about
how binary evolution can give explanation to the observational CMDs
and SEDs of stellar populations.

\section{Binary Evolution Explanation to CMDs}
CMD is well known as one of the best information desk for studying
stars, star clusters, galaxies and the universe. I tried to explain
the CMDs of stellar populations via binary evolution in my work.
According to the observational results, I aims to reproduce the
observational shape and blue stragglers in CMDs. In this work, I
assume that all stars in a simple stellar population form in a star
burst, and stars in a composite stellar population form in different
star bursts. The masses of stars are generated randomly via some
widely used initial mass functions. Then all stars are evolved by a
rapid star evolution code of Hurley et~al. (2002). Finally, the
stellar evolution parameters of stars are transformed into CMDs of
stellar populations via a BaSeL library (Westera et~al. 2002). Some
results can be seen in Figures 2, 3, and 4.

Figure 2 shows a $V$ versus $(B-V)$ CMD of a simple stellar
population including 50\% binary stars with orbital period less than
100 days. The typical observational uncertainty in $B$ and $V$
magnitudes have been taken into account via an uniform distribution.
We find that the shape of the CMD is close to the observational ones
of star clusters, and binary evolution generate some blue stragglers
naturally. This suggests that binary evolution is much better for
CMD studies compared to single evolution.  Binary evolution should
be taken into account all astrophysical studies respecting to stars.

For a comparison purpose, a CMD of a simple stellar population
consisting of single stars is also shown in Figure 3. It shows that
blue stragglers can not be formed via single stellar evolution. In
addition, there are clear differences between the CMDs of simple
stellar populations basing on binary and single stellar evolution.

In Figure 4, a CMD of a composite stellar population with 50\%
binary stars is shown. It consumes that stars in the population have
the same metallicity but different ages. We see that the CMD is very
complicated. If it is compared to the CMDs of some galaxies, we will
find that the CMD shown here is similar to the observational CMDs of
some galaxies. Note that in order to make it easier to plot the CMD,
observational uncertainties are not taken into account in this
Figure. In fact, observational uncertainty has obvious effect on
modeling CMDs.

One can refer to one of our previous work (Li et~al. 2010) to see
how we can use stellar population model basing on binary evolution
for CMD studies of star clusters.

\begin{figure} 
  \includegraphics[angle=-90,width=0.9\textwidth]{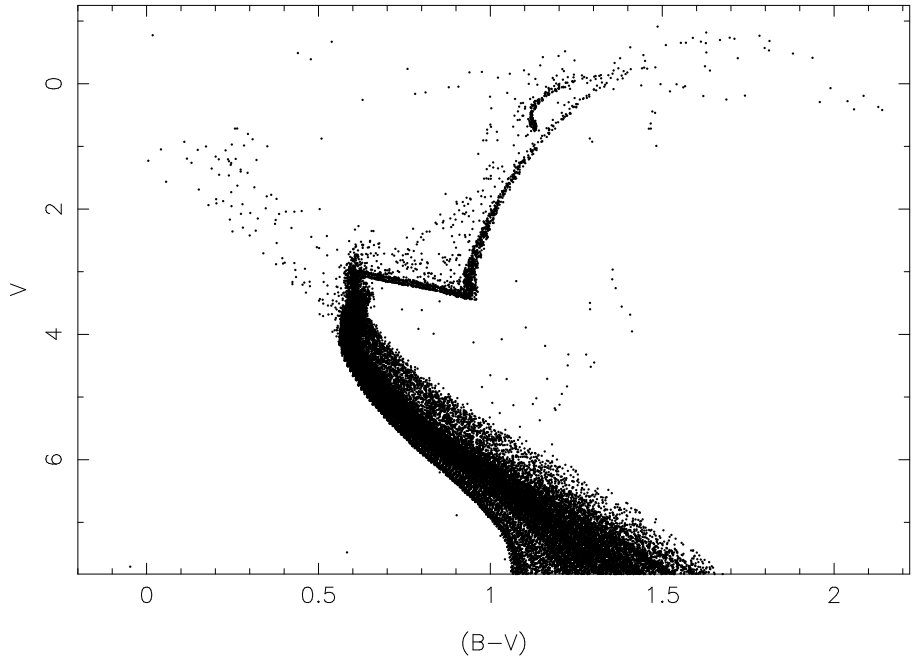}
  \caption{Theoretical CMD of a simple stellar population (50\,000 stars) with 50\% binary stars.
   The metallicity ($Z$) and age of the population are 0.02 and 4.6\,Gyr, respectively.
  }
\end{figure}

\begin{figure} 
  \includegraphics[angle=-90,width=0.9\textwidth]{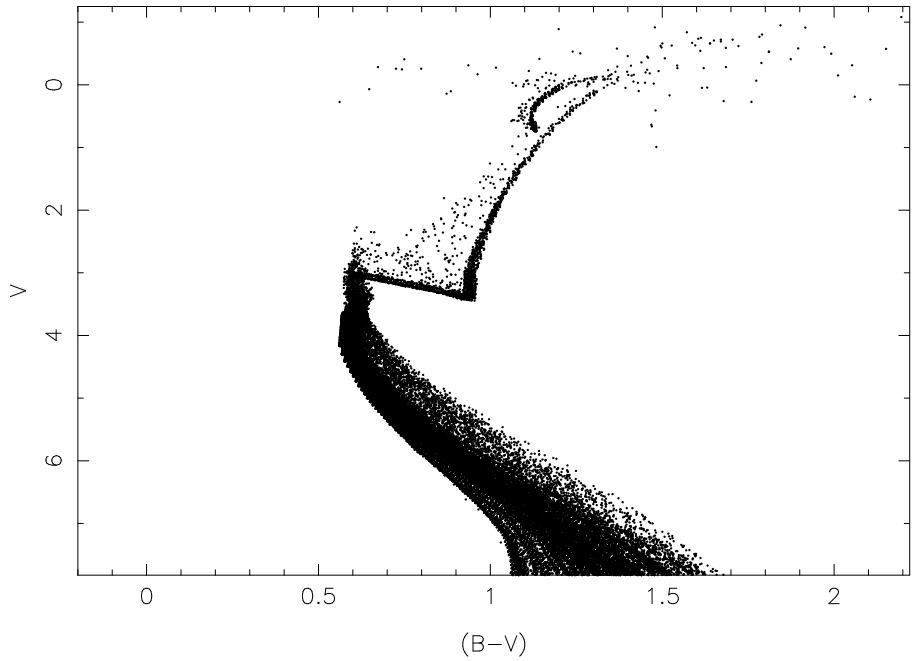}
  \caption{Similar to Figure 2, but for a stellar population without binary stars.}
\end{figure}

\begin{figure} 
  \includegraphics[angle=-90,width=0.9\textwidth]{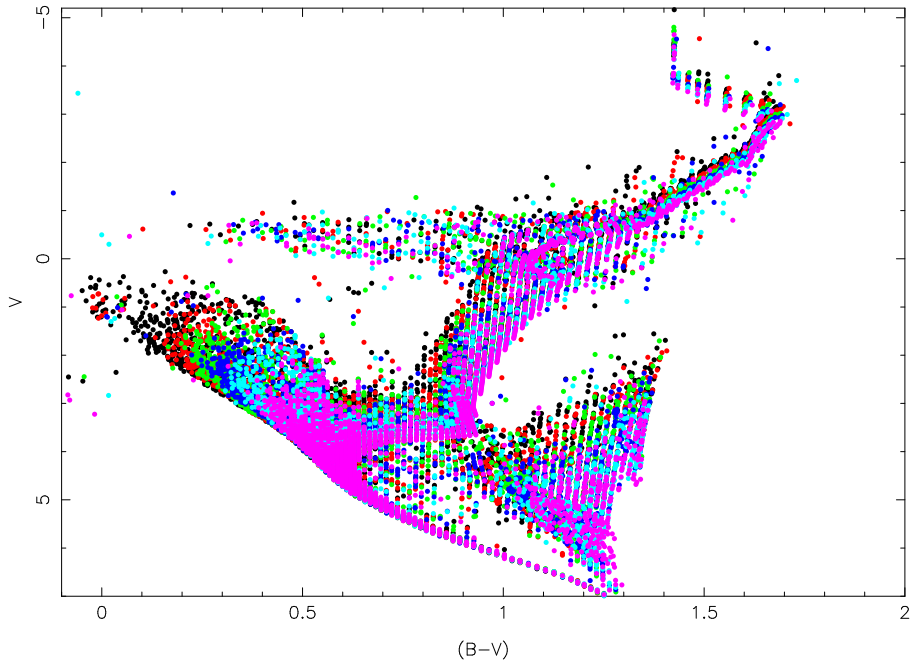}
  \caption{CMD of a composite stellar population with 50\% binary stars.
  Stars are assumed to form between 10 and 5\,Gyr.
  The stellar metallicity ($Z$) of the population is 0.02. Effects of observational uncertainty is not taken into account.}
\end{figure}

\section{Binary Evolution Explanation to Stellar Population SEDs}
Besides CMDs, SEDs are of special importance for astrophysical
studies. SEDs can be used for studying the element abundances,
stellar ages, distances and many other astrophysical properties of
objects. Here I show how binary evolution can give explanation to
SEDs of stellar populations. I pay attention to some UV-upturn SEDs,
because such SEDs are observed in most of elliptical galaxies and it
seems difficult to be explained via single stellar evolution under
nature assumptions. Figure 4 shows the results. We can see that
UV-upturn SEDs are generated naturally by binary evolution for some
stellar populations (see also Han et~al. 2007). Therefore, binary
evolution are possibly the main reason for causing UV-upturn SEDs.
In addition, the figure shows that the age of young component
affects SEDs significantly.

\begin{figure} 
  \includegraphics[angle=-90,width=0.9\textwidth]{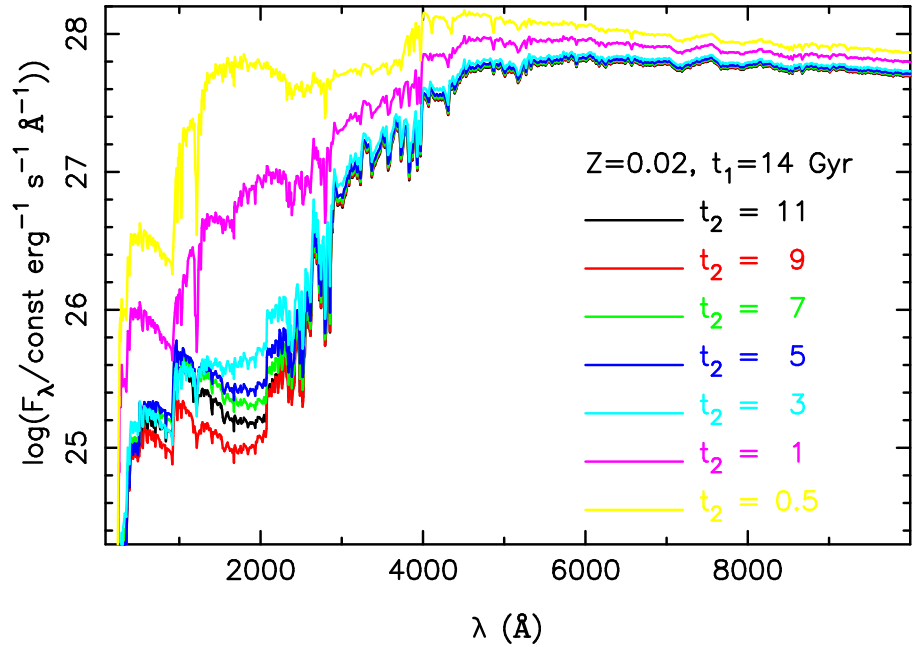}
  \caption{SEDs of a few composite stellar populations with solar stellar metallicity.
   Each composite stellar population contains an old and a young component.
   All populations have the same age (14\,Gyr) for their old components but different ages for their young components.}
\end{figure}

\section{Discussion}
Although this paper discusses the application of binary evolution in
studies of stars, star clusters and galaxies, it needs deeper
investigations. First, all stellar populations take a binary
fraction of 50\%, but the fractions may be different for various
objects. Thus it is necessary to investigate how binary fraction
affect astrophysical studies and build stellar population models
with wide binary fractions. Second, the effect of mass ratio of two
components of a binary should be studied according to observational
results. Third, the metallicity difference in stars of a stellar
population should be noted. This will be important for objects that
have long star formation histories. Finally, a lot of processes in
binary evolution remains unclear. Thus the uncertainty in binary
evolution should be taken into account.

\acknowledgements The author would like to the thank Profs. ZHAO
Gang and HAN Zhanwen for useful suggestions. I also have to thank
Chinese National Science Foundation (Grant No. 10963001), Yunnan
Science Foundation (No. 2009CD093), Chinese Postdoctoral Science
Foundation, Sino-German Center (GZ585) and K. C. Wong  Education
Foundation Hong Kong for their kindly support to my research works.

\end{document}